\documentclass[twocolumn,aps,prl,bm,amsmath,superscriptaddress]{revtex4-1}
\usepackage{amssymb}
\usepackage{mathrsfs}
\usepackage{graphicx,graphics,subfigure}
\usepackage{epstopdf}
\usepackage{epsfig}
\usepackage{float}
\usepackage{txfonts}
\usepackage{xcolor}
\usepackage{siunitx}
\begin{document}
\title{Probabilistic quantum cloning of $N$ quantum states}

\author{Haixin Liu}
\affiliation{School of
	Physics, Peking University, Beijing 100190, China}

\author{Heng Fan}
\email{hfan@iphy.ac.cn}
\affiliation{Beijing National Laboratory for Condensed Matter Physics, Institute of
Physics, Chinese Academy of Sciences, Beijing 100190, China}
\affiliation{CAS Center for Excellence in Topological Quantum Computation,
School of Physical Sciences, University of Chinese Academy of Sciences, Beijing 100190, China}
\affiliation{Beijing Academy of Quantum Information Sciences, Beijing 100190, China}
\affiliation{Songshan Lake Materials Laboratory, Dongguan 523808, China}

\date{ \today}

\begin{abstract}
 Probabilistically creating $n$ perfect clones from $m$ copies for one of $N$ priori known quantum states
  with minimum failure probability is a long-standing problem. We provide a rigorous proof for the geometric approach to this
  probabilistic quantum cloning problem when $N=2$. Then, we give the general geometric form of the sufficient and necessary condition of
  probabilistic cloning for $N$ known quantum states. By this general geometric approach,
  we realize the optimal probabilistic quantum cloning of $N$ known quantum states with priori probabilities.
  The results are also applicable to the identification of those $N$ quantum states.
\end{abstract}

\maketitle
\emph{Introduction.}---No-cloning theorem is fundamental in quantum mechanics and quantum information science \cite{no-cloning,no-cloning-Dieks}.
It states that an unknown quantum state cannot be cloned perfectly. On the other hand, no-cloning theorem
does not forbid imperfect cloning or perfect cloning with probabilistic success.
In the past years, various quantum cloning schemes and related topics
have been studied and implemented experimentally \cite{Ge,YLZhang,ZCYang,teleclone,Derka-Buzek-Ekert,phase-estimation,Chiribella1,DuJF,BuzekHillery,HFanequatorial,GisinMassar,HFanuniversal,
HFandphase,NV-experiment,WangYN,BrussEkertMacchiavello,Wilde,Duexperiment,Science}, see reviews \cite{RMP-cloningReview,HF-report}.
The well studied imperfect quantum cloning machines include, for examples, universal quantum cloning machines and phase-covariant
quantum cloning machines. The aim is to achieve the optimal fidelity for the clones created.
The explicit cloning transformations are given, thus in principle can be directly implemented.

The probabilistic quantum cloning was proposed and studied by Duan and Guo \cite{Id}. The figure of merit is
the success probability in creating perfect clones, i.e., with minimum failure probability.
It is shown that the explicit form of the cloning machine depends on solving a series of inequalities given by positive semidefinite
condition. It is generally a hard task to find the solution to those inequalities, except the simplest case of cloning
probabilistically two states, hereafter, restriction of linear independent is assumed for the states being cloned.
Recently, a geometric approach (GA) was presented in studying probabilistic cloning two states, $N=2$,
in a more general case of creating $n$ copies from $m$ identical input states with priori probabilities.
However, the general $N$ states approach is still unknown, because this GA cannot be generalized straightforwardly.
In this Letter, first we will provide a rigorous proof for GA of probabilistic cloning in $N=2$, resulting from an equality based on
the cloning condition. Then, we present the general GA approach for the probabilistic cloning of $N$ states with
priori probabilities.

\emph{The general framework of probabilistic cloning.}---
A probabilistic cloning machine of $N$ quantum states has an
input port, an output port, and two kinds of flags which we can know the
success or failure of cloning by measuring. A cloning machine can be represented
by a unitary transformation with the help of an ancillary space,
see for example \cite{Id},
\begin{equation}
\begin{split}
U\left| \Psi_{i}\right\rangle^{\otimes m}\otimes\left|0\right\rangle=&\sqrt{p_{i}} \left|\Psi_{i}\right\rangle^{\otimes n}\otimes\left|\alpha_i\right\rangle+\sqrt{q_{i}}\left|\Phi_{i}\right\rangle\otimes\left|\beta_{i}\right\rangle, \\i=&1,\dots,N.
\end{split}
\label{form}
\end{equation}
$\left| \Psi_{i} \right\rangle ,i = ,1, \dots ,N $ are the $N$ known states in a Hilbert space $H$. The machine supplements the input $\left| \Psi_{i} \right\rangle ^{\otimes m}$ with an initial
ancilla $\left|0\right\rangle$ in the ancillary space $H^{\otimes (n-m)} \otimes H_{F}$.
The space $H^{\otimes (n-m)}$ accommodates the additional $n-m$ clones, and $H_{F}$ accommodates the success or failure flags. Set $H_{suc}$ to be the subspace of $H_{F}$ with success flags, and $H_{fail}$ to be the subspace of $H_{F}$ with failure flags. So, $H_{suc}$ is orthogonal to $H_{fail}$, and $H_{F} = H_{suc} \oplus H_{fail}$.
The operator $U$ is the unitary transformation applying to $ H ^{\otimes m} \otimes H ^{\otimes (n-m)} \otimes H_{F}$.
We denote $\left| \alpha_{i}\right\rangle \in H_{suc}$ as the success flags while $\left| \beta_{i}\right\rangle \in H_{fail}$ as failure flags, and $\left|\Phi_{i}\right\rangle$ are failure results in $H^{\otimes n}$. Simply, we shall rewrite  $\left|\Phi_{i}\right\rangle\otimes\left|\beta_{i}\right\rangle$ as $\left| fail_{i} \right\rangle \in H ^{\otimes n} \otimes H_{fail}$ in the rest of this article.
After measuring flags, we obtain $n$ clones and $\left| \alpha_{i}\right\rangle$ if the result is in $H_{suc}$,
and obtain $\left| fail_{i} \right\rangle$ if the result is in $H_{fail}$. Therefore, $ q_{i} $ is the probability of failure when the input state is $\left| \Psi_{i} \right\rangle ^{\otimes m}$. According to the orthogonality, we know $ p_{i} + q_{i} = 1, 0\le p_{i}, q_{i}\le 1$.

Since we do not know which one of the $N$ states is input, we have to assume, from a Bayesian viewpoint, that the state $\left| \Psi_{i} \right\rangle ^{\otimes m}$ is input with some prior probability $ \eta _{i} $, $\sum _{i=1}^{N}\eta _{i}=1$. Then the average probability that the cloner will fail in cloning is $ Q = \sum _{i=1}^{N}\eta _{i} q_{i} $.

\emph{Known results of probabilistic cloning.}---
Due to the unitarity of $U$, $ q_{i} $, $\left| \alpha_{i}\right\rangle $ and $\left| fail_{i} \right\rangle $ cannot be selected arbitrarily. Here we introduce the notations,
$ X ^{(r)} = \left[ \left\langle \Psi_{i}|\Psi_{j} \right\rangle ^{r}\right]  _{ij} $,
$ X ^{(r)} _{p} = \left[ \left\langle \Psi_{i}|\Psi_{j} \right\rangle ^{r} \left\langle \alpha_{i} | \alpha_{j} \right\rangle \right] _{ij} $,
$ \Gamma = {\rm diag.} (p_{1},\dots ,p_{N}) $,
$\sqrt{\Gamma}={\rm diag.} (\sqrt{p_{1}},\dots ,\sqrt{p_{N}})$,
$ Y = \left[ \left\langle fail_{i} |fail_{j}\right\rangle \right] _{ij} $, where
$X ^{(r)}, X ^{(r)} _{p}, \Gamma, \sqrt{\Gamma}, Y$ are five $N\times N$ matrices.

In Ref.\cite{Id}, two theorems of probabilistic cloning are presented.

\emph{Theorem 1}: That $U$  is unitary is equivalent to
     \begin{equation}
      \begin{split}
      X ^{(m)} = \sqrt{\Gamma} X ^{(n)} _{p} \sqrt{\Gamma}+ \sqrt{ I_{N}-\Gamma} Y \sqrt{ I_{N}-\Gamma}
      \end{split}
      \label{uni}
     \end{equation}
$I_N$ is a $N$-dimensional identity matrix.

{\it Theorem 2}: The probabilistic cloning of $N$ states can be realized with the parameter matrix $\Gamma$ if and only if the matrix $ X ^{(m)} - \sqrt{\Gamma} X ^{(n)} _{p} \sqrt{\Gamma} $ is positive semidefinite.

However, in order to know whether this matrix is positive semidefinite,
we have to solve a series of inequalities \cite{Id}, which is a hard task.
Besides, the general case that these $N$ states have arbitrary priori probabilities has not yet been investigated.

Recently, the GA is presented to the  optimization of the probabilistic cloning of two states \cite{Ge}. Let us explain the results and
some notations,
$s_{ij}=\left\langle \Psi_{i} | \Psi_{j} \right\rangle$,
$\alpha_{ij}=\left\langle \alpha_{i}| \alpha_{j}\right\rangle$,
$f_{ij}=\left\langle fail_{i}| fail_{j}\right\rangle$,
$i,j \in \left\lbrace 1,\dots,N\right\rbrace$. The subscript will be omitted when $N=2$ and $(i,j)=(1,2)$.
Instead of solving inequalities, an equality is proposed to determine the optimal probabilistic cloning,
	\begin{equation}
	\begin{split}
	 s ^ {m} = \sqrt{p_{1}p_{2}} s ^ {n} \alpha + \sqrt{q_{1}q_{2}}.
	\end{split}
	\label{N=2}
	\end{equation}
 This equation represents a class of curves with different $\alpha$ in the parameter space, $0\le q_{1}, q_{2}\le 1$. Moveover, with the known priori probabilities $ \eta =(\eta_{1} ,\eta_{2})$, the average failure probability of a point $ J=(q_{1} ,q_{2}) $ in the parameter space is proportional to the distance between the origin point and the line which is orthogonal to vector $\vec{\eta}$ and $J$ is located on. In order to obtain this equality, it is argued in Ref \cite{Ge} that an optimal cloning requires $|f|=1$, and $s$ and $\alpha $ can be regarded as nonnegative real numbers without any loss of generality. Due to the limited space, we will show our rigorous proof of $|f|=1$ and equality (\ref{N=2}) in the Appendix.

\emph{The geometric forms of sufficient and necessary condition of probabilistic cloning and a necessary equality for optimization}---
Similar to the case of two states, in order to apply GA to the case of N states, we have to construct, in the parameter space $V$: $0\le q_{i}\le 1, i=1,\dots ,N$, a $N-1$-dimensional surface on which all parameter points of optimal probabilistic cloning are. In the case of two states, with the condition $|f|=1$, the equation of this surface can be obtained from the inner product of formula (\ref{form}) with different i. However, for the optimal probabilistic cloning of three or more states, the condition $\left| \left\langle fail_{1}| fail_{2}\right\rangle\right| = 1$ cannot always be satisfied, as an example of three states in the Appendix shows. Therefore, we turn to make use of Theorem 2 to construct the surface by proving  Theorem 3.

{\it Theorem 3}: A necessary condition for the optimal probabilistic cloning of N known states is
:
	\begin{equation}
	det( X ^{(m)} - \sqrt{\Gamma} X ^{(n)} _{p} \sqrt{\Gamma}) = 0
	\label{nec}
	\end{equation}
Before proving Theorem 3, let's see a useful lemma:

{\it Lemma 1}: The determinant of a N-dimensional positive semidefinite Hermitian matrix equals to 0, if one of its principal minors equals 0.

We will use mathematical induction to prove this theorem.

The proposition is right when $N=1$.

Supposing that the proposition is right when $N\le K, K$ is a positive integer, let's see the case of $N=K+1$. Set this matrix to be $M_{K+1}$ and one of its principal minors which equals 0 to be $D$. (1)If the dimension of $D$ is $K+1$, the proposition will be right when $N=K+1$. (2)If the dimension of D is smaller than $K+1$. There exists a K-dimensional principal submatrix $M_{K}$ and $D$ is also its principal minor. According to the definition of principal submatrix, $M_{K}$ is also a positive semidefinite Hermitian matrix. According to the Inductive hypothesis, the determinant of $M_{K}$ is 0. Without any loss of generality, we can rewrite $M_{K+1}$ as
\begin{equation}
M_{K+1}= \begin{pmatrix}
M_{K} & b \\
b^ {\dag} & d
\end{pmatrix}
\end{equation}
$d$ is real number and $b$ is a $K$-dimensional column vector. Since the determinant of $M_{K}$ is 0, there exists such a $K$-dimensional nonzero column vector $x_{1}$ that $M_{K} x_{1}=0$. Notice that for any complex number $x_2$,
$x = \begin{pmatrix}
	x_{1} \\
	x_{2}
\end{pmatrix}$ is a nonzero $K+1$-dimensional column vector. Since $M_{K+1}$ is positive semidefinite,
\begin{equation}
\begin{split}
x^ {\dag}M_{K+1}x&=(x^ {\dag}_{1},x^ {*}_{2})
\begin{pmatrix}
M_{K} & b \\
b^ {\dag} & d
\end{pmatrix}
\begin{pmatrix}
x_{1} \\
x_{2}
\end{pmatrix}\\
& =x^ {\dag}_{1}M_{K} x_{1}+ x_{2}x^ {\dag}_{1}b+x^ {*}_{2} b^ {\dag}x_{1}+d\left| x_{2}\right|^ {2} \\
& =2Re(x_{2}x^ {\dag}_{1}b) +d\left| x_{2}\right|^ {2}\ge 0
\end{split}
\label{M}
\end{equation}
Rewrite $x_2$ as $x_{2} =e^ {i\delta}r, r\ge 0$, $\delta$ and $r$ are real numbers. Since $x_2$ is an arbitrary complex number, we can choose $\delta$ to satisfy $ 2Re(x^ {*}_{2} b^ {\dag}x_{1})=-2r\left| b^ {\dag}x_{1}\right|$. Then, inequality (\ref{M}) can be rewritten as
\begin{equation}
dr^ {2}\ge 2r\left| b^ {\dag}x_{1}\right| .
\end{equation}
According to the arbitrariness of $x_2$ and the inequality above, for any $r>0$, $0\le \left| b^ {\dag}x_{1}\right|\le \frac{dr}{2}$. Therefore,
\begin{equation}
b^ {\dag}x_{1}=0
\end{equation}
Considering a nonzero $K+1$ dimensional column vector $ x^ {'} = \begin{pmatrix}
x_{1} \\
0
\end{pmatrix}$,
\begin{equation}
M_{K+1}x= \begin{pmatrix}
M_{K} & b \\
b^ {\dag} & d
\end{pmatrix}
\begin{pmatrix}
x_{1} \\
0
\end{pmatrix} =\begin{pmatrix}
M_{K} x_{1} \\
b^ {\dag}x_{1}
\end{pmatrix} = 0
\end{equation}
Hence, the determinant of $M_{K+1}$ equals 0. Therefore, in both cases, the proposition is right when $N\le K+1$.

According to mathematical induction, the proposition is right for any $ N\ge 1$.

Now, let us prove Theorem 3.

Proof: Suppose that a cloning is optimal and its parameter point in the parameter space $V$ is $q=(q_{1} ,\dots ,q_{N})$. According to Theorem 2, $ X ^{(m)} - \sqrt{\Gamma} X ^{(n)} _{p} \sqrt{\Gamma} $ is positive semidefinite. Since this matrix is an Hermitian matrix, according to the knowledge of linear algebra, it is positive semidefinite if and only if all its principal minors are nonnegative.

Supposing that all principal minors of this matrix are positive at point $q$, due to the principal minors' continuity related to $(q_{1} ,\dots ,q_{N})$, we know that there exists such $ (q^ {'}_{1} ,\dots ,q^ {'}_{N}) $ that for all $i \in \left\lbrace 1,\dots,N\right\rbrace ,q^ {'}_{i} <q_{i}$ and all minors of the matrix$ X ^{(m)} - \sqrt{\Gamma} X ^{(n)} _{p} \sqrt{\Gamma} $ are still positive, which means that probability cloning can be realized under the new parameter. As a result, for any $ \eta =(\eta_{1} ,\dots ,\eta_{N})$, the new failure probability $ Q^ {'} = \sum _{i=1}^{N}\eta _{i} q^ {'}_{1} <Q$, which contradicts that the previous probabilistic cloning is optimal. Therefore, in order to achieve optimal probabilistic cloning, at least one of principal minors of $ X ^{(m)} - \sqrt{\Gamma} X ^{(n)} _{p} \sqrt{\Gamma} $ equals 0. According to Lemma 1,
\begin{equation}
det( X ^{(m)} - \sqrt{\Gamma} X ^{(n)} _{p} \sqrt{\Gamma}) = 0
\end{equation}
Then we complete our proof.

In order to apply GA, a necessary condition alone is not enough. Before we give the geometric form of sufficient and necessary condition of probabilistic cloning of N known states, we would like to introduce what equality (\ref{nec}) means in the parameter space $V$. $det( X ^{(m)} - \sqrt{\Gamma} X ^{(n)} _{p} \sqrt{\Gamma}) = 0$ determines one or several pieces of $N-1$-dimensional surfaces in $V$. These surfaces divide the rest of the space $R$: $det( X ^{(m)} - \sqrt{\Gamma} X ^{(n)} _{p} \sqrt{\Gamma}) \ne 0$ into different connected sets. We will prove an important conclusion first:

{\it Lemma 2}: Probabilistic cloning can be realized at all the parameter points in a connected subset of $R$ if and only if it can be realized at a parameter point in this connected subset.

Proof:
If probabilistic cloning can be realized at all the parameter points in a connected subset in $R$, it can be realized at a parameter point in this connected subset.

If probabilistic cloning can be realized at a parameter point $J$ in a connected set $W\subset R$, $X ^{(m)} - \sqrt{\Gamma} X ^{(n)} _{p} \sqrt{\Gamma}$ are positive semidefinite. Due to the definition of R, $det( X ^{(m)} - \sqrt{\Gamma} X ^{(n)} _{p} \sqrt{\Gamma}) \ne 0$ at point $J$. According to Lemma 1, all principal minors of $X ^{(m)} - \sqrt{\Gamma} X ^{(n)} _{p} \sqrt{\Gamma}$ are positive at $J$.
Let's see what will happen to an arbitrary point $Z$ in the set $W$. According to the definition of connected set, there exists a continuous curve $C$ in $W$ which connects point $J$ to point $Z$. Set the parameter of this curve to be $t\in[0,1]$, $t=0$ at point $J$ and $t=1$ at point $Z$. Supposing that probabilistic cloning cannot be realized at point $Z$, some of principal minors of $X ^{(m)} - \sqrt{\Gamma} X ^{(n)} _{p} \sqrt{\Gamma}$ are negative at $Z$. Thus, there exist some of principal minors that equal 0 at some points of curve $C$. We note all this kind of principal minors as $D_j, j=1,\dots,M$. Due to $D_j$'s continuity related to $ (q^ {'}_{1} ,\dots ,q^ {'}_{N}) $ and the continuity of the curve $C$, $D_j$ can be regarded as a continuous function of $t$. For each $D_j$, according to the knowledge of Mathematical Analysis, there exists a minimum value $0<t_j<1$ at which $D_j=0$. Set the minimum value of all $t_j$ to be $t_{min}$, which represents a point $G$ on curve $C$. As a result, at point $G$, some of the principal minors of $X ^{(m)} - \sqrt{\Gamma} X ^{(n)} _{p} \sqrt{\Gamma}$ equal 0 and all of the principal minors of $X ^{(m)} - \sqrt{\Gamma} X ^{(n)} _{p} \sqrt{\Gamma}$ are non-negative, which means that $X ^{(m)} - \sqrt{\Gamma} X ^{(n)} _{p} \sqrt{\Gamma}$ is positive semidefinite. According to Lemma 1, $det( X ^{(m)} - \sqrt{\Gamma} X ^{(n)} _{p} \sqrt{\Gamma}) = 0$, which contradicts the definition of $R$. Therefore, probabilistic cloning can be realized at point $Z$.

Then we complete our proof.

There might be several connected subsets in R. In order to exclude some of the connected subsets in R, we make use of the specific form of the matrix $X ^{(m)} - \sqrt{\Gamma} X ^{(n)} _{p} \sqrt{\Gamma}$ and get another Lemma:

{\it Lemma 3}: If probabilistic cloning can be realized at a point $J$, the matrix $X ^{(m)} - \sqrt{\Gamma} X ^{(n)} _{p} \sqrt{\Gamma}$ is positive definite at point $(1,\dots,1)$ and all points on the segment between $J$ and point $(1,\dots,1)$.

Proof:
According to Theorem 2, if probabilistic cloning can be realized at a point $J=(q_1,\dots,q_N)$, $X ^{(m)} - \sqrt{\Gamma} X ^{(n)} _{p} \sqrt{\Gamma}$ is positive semidefinite at $J$. Besides, according to Ref \cite{Id}, the input states $\left|\Psi_{i}\right\rangle^{\otimes m},i=1,\dots,N$ are linenar independent. Therefore, for any nonzero vector $x^T=(x_1,\dots,x_N)^T $,
\begin{equation}\begin{split}
x^{\dag}\sqrt{\Gamma} X ^{(n)} _{p} \sqrt{\Gamma}x
&=\left\| \sum_{i=1}^{N}x_i\sqrt{p_{i}} \left|\Psi_{i}\right\rangle^{\otimes n}\otimes\left|\alpha_i\right\rangle\right\|^{2}\ge 0\\
x^{\dag}X ^{(m)}x
&=\left\| \sum_{i=1}^{N}x_i\left|\Psi_{i}\right\rangle^{\otimes m}\right\|^{2}>0
\end{split}\end{equation}
which means the matrix $\sqrt{\Gamma} X ^{(n)} _{p} \sqrt{\Gamma}$ is always positive semidefinite and the matrix  $X ^{(m)}$ is positive definite. As a result, for any real number $0\le r<1$ and any nonzero vector $x^T=(x_1,\dots,x_N)^T $,  if $x^{\dag}\sqrt{\Gamma} X ^{(n)} _{p} \sqrt{\Gamma}x\ne0$
\begin{equation}\begin{split}
x^{\dag}(X ^{(m)} - \sqrt{r\Gamma} X ^{(n)} _{p} \sqrt{r\Gamma})x
>x^{\dag}(X ^{(m)} - \sqrt{\Gamma} X ^{(n)} _{p} \sqrt{\Gamma})x\ge0
\end{split}\end{equation}
and if $x^{\dag}\sqrt{\Gamma} X ^{(n)} _{p} \sqrt{\Gamma}x=0$
\begin{equation}\begin{split}
x^{\dag}(X ^{(m)} - \sqrt{r\Gamma} X ^{(n)} _{p} \sqrt{r\Gamma})x=x^{\dag}X ^{(m)}x>0
\end{split}\end{equation}
Hence,for any real number $0\le r<1$,$X ^{(m)} - \sqrt{r\Gamma} X ^{(n)} _{p} \sqrt{r\Gamma}$ will be positive definite at $J$ , which means that $X ^{(m)} - \sqrt{\Gamma} X ^{(n)} _{p} \sqrt{\Gamma}$ is also positive definite at the point $ (1-r(1-q_1),\dots,1-r(1-q_N))$.

Then we complete our proof.

Besides, combining Lemma 3 with Theorem 2,we can obtain that if probabilistic cloning can be realized at a point $J$, it can be realized at any point on the segment connecting $J$ and point $(1,\dots,1)$.

Furthermore,according to Lemma 3, if $J$ is in $R$, which means $X ^{(m)} - \sqrt{\Gamma} X ^{(n)} _{p} \sqrt{\Gamma}$ is positive definite at $J$, $X ^{(m)} - \sqrt{\Gamma} X ^{(n)} _{p} \sqrt{\Gamma}$ will be positive definite at all points on the segment connecting $J$ and point $(1,\dots,1)$. As a result, this segment $J$ and point $(1,\dots,1)$ is in $R$, which means $J$ is in the same connected subset with point $(1,\dots,1)$.Hence, we exclude all connected subsets in R except one. We use $B$ to represent this subset and get the following conclusion:

{\it Lemma 4}: All of points in $R$ where probabilistic cloning can be realized are in the same connected subset with point $(1,\dots,1)$.

In terms of those points on the surfaces determined by $det( X ^{(m)} - \sqrt{\Gamma} X ^{(n)} _{p} \sqrt{\Gamma}) = 0$, we can know whether probabilistic cloning can be realized at them by using the following conclusion:

{\it Lemma 5}: Probabilistic cloning can be realized at point $J$ where $det( X ^{(m)} - \sqrt{\Gamma} X ^{(n)} _{p} \sqrt{\Gamma}) = 0$ if and only if $J$ is a boundary point of $B$.

Proof:
 If a parameter point $J$ is a boundary point of $B$, it is a limit point of $B$ or a point of $B$. Due to the continuity of the principal minors of $X ^{(m)} - \sqrt{\Gamma} X ^{(n)} _{p} \sqrt{\Gamma}$ related to $ (q^ {'}_{1} ,\dots ,q^ {'}_{N}) $, all the principal minors of $X ^{(m)} - \sqrt{\Gamma} X ^{(n)} _{p} \sqrt{\Gamma}$ are non-negative at point $J$. Therefore, $X ^{(m)} - \sqrt{\Gamma} X ^{(n)} _{p} \sqrt{\Gamma}$ is positive semidefinite at $J$ and probabilistic cloning can be realized at $J$.

 On the other hand,according to Lemma 3,since probabilistic cloning can be realized at $J$, probabilistic cloning can be realized at all the points on the segment connecting $J$ and point $(1,\dots,1)$, and $det( X ^{(m)} - \sqrt{\Gamma} X ^{(n)} _{p} \sqrt{\Gamma}) > 0$ at these points except $J$. Thus, the points on this segment except $J$ are in set $B$. Therefore, $J$ is a boundary point of $B$.

Then, combining Lemma 4 and Lemma 5, we can give the sufficient and necessary condition of probabilistic cloning of N known states in a geometric form:

{\it Theorem 4}: Probabilistic cloning can be realized at and only at the closure of set $B$. $B$ is a connected subset of the set where the matrix $X ^{(m)} - \sqrt{\Gamma} X ^{(n)} _{p} \sqrt{\Gamma}$ is non-degenerate, and includes the point $q_i=1, i=1,\dots,N $

Besides, Theorem 4 helps us exclude all surfaces determined by Theorem 3 except the boundary of $B$. We will note this boundary as $S$ and rewrite Theorem 3 in a stronger form:

{\it Theorem 3*}: All the points where the probabilistic cloning is optimal are on $S$.

Theorem 3* and Theorem 4 means that we only need to solve one equality and do not have to solve any inequality at all.

\emph{The generalization of GA in the case of N states.}---
Now, we will introduce our GA. For any given $ \eta =(\eta_{1} ,\dots ,\eta_{N})$, the average probability of a failure cloning at a point $ J=(q_{1} ,\dots ,q_{N}) $ in the parameter space $V$ is the $ \sqrt{ \sum _{i=1}^{N}\eta _{i} ^ {2}} $ times of the distance between the origin point and the plane which is orthogonal to vector $ \vec{\eta} =(\eta_{1} ,\dots ,\eta_{N}) $ and  $ J=(q_{1} ,\dots ,q_{N}) $ is on. Therefore, if $J$ is optimal probabilistic cloning, the plane should be a tangent of S or touch the margin of S at a point when some of the $q_i$ equal 1.

Although our GA can only realize the optimization of probabilistic cloning of N states when $\alpha_{ij},i,j=1,\dots,N$ is given and how to select $\alpha_{ij},i,j=1,\dots,N$ to optimize probabilistic cloning is still a problem when $N\ge 3$,  the probabilistic identification of N states is not related to $\alpha_{ij},i,j=1,\dots,N$. According to Ref \cite{Id}, the probability of failure to identify one state of N known states is equal to the probability of failure to clone an unknown state with the parameter $\alpha_{ij}=\delta_{ij},i,j=1,\dots,N$. Therefore, we can directly obtain the following conclusion:

{\it Corollary 1}: Probabilistic identification can be realized at and only at the closure of set $B$.$B$ is a connected subset of the set where the matrix $X ^{(1)} - \Gamma$ is non-degenerate, and includes the point $q_i=1, i=1,\dots,N $.

{\it Corollary 2}: A necessary condition of optimal probabilistic identification is that $det(X ^{(1)} - \Gamma)=0$.

Accordingly, for any given $ \eta =(\eta_{1} ,\dots ,\eta_{N})$, we can apply our GA to the optimization of the identification of N states.

\begin{figure}[htbp]
  \centering
  \includegraphics[width=0.5\linewidth]{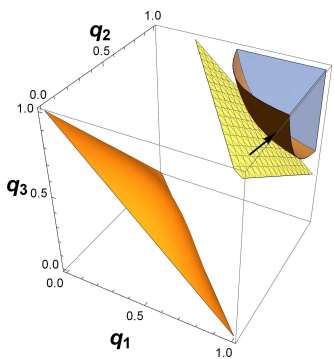}\\
  \caption{The cube in this figure represents the parameter space V. Two seperate organe surfaces are determined by Equality (\ref{num}). The blue region represents the set $B$. The organe surface at the right top corner represents the boundary of $B$. The black arrow is parallel to $ \vec{\eta} $. The yellow plane with mesh  which is orthogonal to the black arrow is an auxiliary plane which helps us find the optimal probabilistic cloning with the given $ \vec{\eta} $.}
  \label{Fig1}
\end{figure}

\emph{An example of our GA when $N=3$}---
In order to make it clear, we will show this approach in the following examples.Firstly, let $N=3$, the matrix can be written as
\begin{equation}
X ^{(1)}-\Gamma=
\begin{pmatrix}
q_{1} & s_{12}& s_{13}\\
s^ {*}_{12} & q_{2}& s_{23}\\
s^ {*}_{13} & s^ {*}_{23} & q_{3}
\end{pmatrix}
\end{equation}
the surfaces can be decided by the following equality.
\begin{equation}
q_{1}q_{2}q_{3}-q_{1}\left| s_{23}\right| ^ {2}-q_{2}\left| s_{31}\right| ^ {2}-q_{3}\left| s_{12}\right| ^ {2}+2Re( s_{12}s_{23}s_{31})=0
\label{ex}
\end{equation}
Take
\begin{equation}
\begin{cases}
& \left| \psi_{1}\right\rangle =\frac{1}{\sqrt{3}}(1,1,1) \\
& \left| \psi_{2}\right\rangle =\frac{1}{\sqrt{3}}(1,1,\omega ) ,\omega=e^{\frac{2\pi}{3}i} \\
& \left| \psi_{3}\right\rangle =\frac{1}{\sqrt{3}}(1,\omega,\omega )
\end{cases}
\end{equation}
as an example. $ s_{12}=s_{23}=\frac{2+\omega}{3}$, $ s_{31}=\frac{1+2\omega^{2}}{3} $
formula (\ref{ex}) can be rewritten as
\begin{equation}
q_{1}q_{2}q_{3}-\frac{1}{3}(q_{1}+q_{2}+q_{3})+\frac{1}{3}=0
\label{num}
\end{equation}
 Equality (\ref{num}) devides the parameter space into different regions with two separate orange surfaces, just as Fig. (\ref{Fig1}) shows.

 Now, we take $\eta=(0.35,0.25,0.4) $ as an example. The small black arrow in Fig.1 is parallel to $ \vec{\eta} $, and the yellow plane with mesh is orthogonal to the black arrow and tangent to the piece of orange surface. The tangent point is the parameters of optimal identification of these states.

 In a  more special case, we still use these states but change $\eta$ into $(\frac{1}{3},\frac{1}{3},\frac{1}{3}) $, which means those three states are put in with the same priori probability. Due to the symmetry of the surface $S$ and the auxiliary plane, we figure out the accurate value of the average probability of success of optimal identification of these three states: $1-\frac{2\cos\ang{50}}{\sqrt{3}}\approx 0.258$. We will show this process in the Appendix.

In summary, we strictly prove the former GA to the optimization of probabilistic cloning of two states in Ref \cite{Ge}. Then, we give the geometric form of the abstract sufficient and necessary condition of the probabilistic cloning of N states (Theorem 4) and a necessary condition of the optimal probabilistic cloning of N states (Theorem 3*). After that, we generalize the GA into the optimization of the cloning with some given parameters and the identification of N known quantum states with Corollary 1 and Corollary 2. Finally, we give two examples of the probabilistic identification of three states to show our approach. Besides, the Lemma 1, Lemma 2 and our GA can also be applied into other optimization problems that involves constraints related to the positive semidefinition of parameter matrices.

\begin{acknowledgments}
	This work was supported by National Key Research and Development Program of China (Grant Nos. 2016YFA0302104, 2016YFA0300600),
	National Natural Science Foundation of China (Grant No.11774406),
and Strategic Priority Research Program of Chinese Academy of Sciences (Grant No. XDB28000000).
\end{acknowledgments}

\section{Appendix}

 In order to obtain equality (\ref{N=2}) that GA relies on, the authors of Ref \cite{Ge} argues that an optimal cloning requires $|f|=1$, and $s$ and $\alpha $ can be regarded as non-negative real numbers without any loss of generality. Here, we will prove these conclusions when $N=2$.

Proof:

We first apply Theorem 2. The probability cloning of two states is realizable if and only if
\begin{equation}
X ^{(m)}-\sqrt{\Gamma} X ^{(n)}_{p} \sqrt{\Gamma} =
\begin{pmatrix}
1-p_{1} & s ^ {m} - \sqrt{p_{1}p_{2}} s ^ {n} \alpha \\
s ^ {*m}-\sqrt{p_{1}p_{2}} s ^ {*n} \alpha^ {*}& 1-p_{2}
\end{pmatrix}
\end{equation}
is positive semidefinite. The matrix above is an Hermitian matrix, so according to the knowledge of linear algebra, it is positive semidefinite if and only if all of its principal minors are nonnegative, that is,
\begin{equation}
\begin{split}
\begin{cases}
& 1-p_{1}\ge 0 \\
& 1-p_{2}\ge 0 \\
& (1-p_{1})( 1-p_{2})-\left| s ^ {m} - \sqrt{p_{1}p_{2}} \alpha s ^ {n} \right| ^2 \ge 0
\end{cases}
\end{split}
\end{equation}
The first and second lines of this equation are trivial and the third line is equivalent to
\begin{equation}
\sqrt{(1-p_{1})( 1-p_{2})}-\left| s ^ {m} - \sqrt{p_{1}p_{2}} \alpha s ^ {n} \right|\ge 0
\end{equation}
If $ \sqrt{(1-p_{1})( 1-p_{2})}-\left| s ^ {m} - \sqrt{p_{1}p_{2}} \alpha s ^ {n} \right| > 0 $, the clone must not be optimal. The reason is that due to the continuity related to $p_{1}, p_{2}$ on the left side, there exist such
$p^{'}_{1}> p_{1}$and$ p^{'}_{2}> p_{2}$ that
$(\sqrt{(1-p^{'}_{1})( 1-p^{'}_{2})}-\left| s ^ {m} - \sqrt{p^{'}_{1}p^{'}_{2}} \alpha s ^ {n} \right|>0)$, which means that probability cloning can be realized under the new parameter $ p^{'}_{1} $ and $ p^{'}_{2} $. For any $ \eta _{1} $ and $ \eta _{2} $, the new average failure probability $ Q^{'} = \eta _{1} (1- p^{'}_{1}) + \eta _{2}(1- p^{'}_{2})<Q $, which means that the previous cloning is not optimal. Hence, for the optimal probabilistic cloning, 	
\begin{equation}\begin{split}
\sqrt{(1-p_{1})( 1-p_{2})}-\left| s ^ {m} - \sqrt{p_{1}p_{2}} \alpha s ^ {n} \right| = 0
\end{split}\end{equation}
Furthermore, we will prove next that if the complex angle of $ s ^ {m} $ is different from that of $ \sqrt{p_{1}p_{2}} \alpha s ^ {n} $, then the probabilistic cloning is not optimal. Notice that
\begin{equation}\begin{split}
\left| s ^ {m} \right| \ge \left| \sqrt{p_{1}p_{2}} \alpha s ^ {n}\right|
\end{split}\end{equation}
If the complex angles of $ s ^ {m} $ and $ \sqrt{p_{1}p_{2}} \alpha s ^ {n} $ are different, we can choose another $ \left| \alpha^ {'}_{2}\right\rangle $ to let $ \alpha^ {'} = \left\langle \alpha^ {'}_{1} |\alpha^ {'}_{2}\right\rangle $ satisfy the same complex angle of $ s ^ {m} $ and $ \sqrt{p_{1}p_{2}} \alpha^ {'} s ^ {n} $ which leads to $\sqrt{(1-p_{1})( 1-p_{2})}-\left| s ^ {m} - \sqrt{p_{1}p_{2}} \alpha^ {'} s ^ {n} \right| > 0$. In this case, with the similar method above, we can reduce Q by selecting a new group of $p_{1}$ and $p_{2} $. Therefore, for the optimal probabilistic cloning, $\sqrt{(1-p_{1})( 1-p_{2})}=\left| s ^ {m} - \sqrt{p_{1}p_{2}} \alpha s ^ {n} \right| = \left| s\right| ^ {m}- \sqrt{p_{1}p_{2}}\left| \alpha\right| \left| s\right| ^ {n}$, which is equivalent to
\begin{equation}
\left| s\right| ^ {m}= \sqrt{p_{1}p_{2}}\left| \alpha\right| \left| s\right| ^ {n} +\sqrt{q_{1}q_{2}}
\label{N=2'}
\end{equation}
Besides, when $N=2$, formula (\ref{uni}) in Theorem 1 is equivalent to
\begin{equation}
s ^ {m} = \sqrt{p_{1}p_{2}} s ^ {n} \alpha + \sqrt{q_{1}q_{2}}\left\langle fail_{1} |fail_{2}\right\rangle
\end{equation}
Combining this formula and formula (\ref{N=2'}), we obtain
\begin{equation}
\sqrt{q_{1}q_{2}}\left| \left\langle fail_{1}| fail_{2}\right\rangle\right| =\left| s\right| ^ {m}- \sqrt{p_{1}p_{2}}\left| \alpha\right| \left| s\right| ^ {n}=\sqrt{q_{1}q_{2}} ,
\end{equation}
If $s=0$, it is obvious that  $q_1=q_2=0$ when the probabilistic cloning is optimal. We can select $|f|=1$ without contradicting the sufficient and necessary condition in Theorem 2. If $0<|s|<1$, $ \left| s ^ {m} \right| > \left| \sqrt{p_{1}p_{2}} \alpha s ^ {n}\right|$, which means $\sqrt{p_{1}p_{2}}>0$. As a result, $|f|=1$. Therefore, in both cases,
\begin{equation}
\left| \left\langle fail_{1}| fail_{2}\right\rangle\right| = 1
\end{equation}
Then, we complete our proof of equality $|f|=1$ and equality (\ref{N=2'}) which is the general form of equality (\ref{N=2}).

However, for the optimal probabilistic cloning of three or more states, the condition $\left| \left\langle fail_{1}| fail_{2}\right\rangle\right| = 1$ cannot always be satisfied. The following group of three linear independent states when $m=1$ are such an example:
\begin{equation}
\begin{cases}
& \left| \Psi_{1} \right\rangle = (1,0,0)\\
& \left| \Psi_{2} \right\rangle = (0,1,0)\\
& \left| \Psi_{3} \right\rangle = (\frac{1}{\sqrt{3}},\frac{1}{\sqrt{3}},\frac{1}{\sqrt{3}})
\end{cases}
\end{equation}
Let us prove that at least one of $f_{ij}$ equals 0. According to formula (\ref{form}), by taking the inner product of this kind of equality with different i, we obtain the following equation:
\begin{equation}
\begin{cases}
& s_{12} = \sqrt{p_{1}p_{2}} s ^ {n}_{12} \alpha_{12} + \sqrt{q_{1}q_{2}} f_{12} \\
&s_{23} = \sqrt{p_{2}p_{3}} s ^ {n}_{23} \alpha_{23} + \sqrt{q_{2}q_{3}} f_{23} \\
& s_{31} = \sqrt{p_{3}p_{1}} s ^ {n}_{31} \alpha_{31} + \sqrt{q_{3}q_{1}} f_{31}
\end{cases}
\end{equation}
In this case, $s_{12}=0, s_{23}=s_{31}=\frac{1}{\sqrt{3}} $. Then we rewrite the equation above as
\begin{equation}
\begin{cases}
& 0 = 0 + \sqrt{q_{1}q_{2}} f_{12} \\
&\frac{1}{\sqrt{3}} = \sqrt{p_{2}p_{3}} (\frac{1}{\sqrt{3}}) ^ {n} \alpha_{23} + \sqrt{q_{2}q_{3}} f_{23} \\
& \frac{1}{\sqrt{3}} = \sqrt{p_{3}p_{1}} (\frac{1}{\sqrt{3}}) ^ {n} \alpha_{31} + \sqrt{q_{3}q_{1}} f_{31}
\end{cases}
\end{equation}
If $f_{12}\ne 0$, according to the first line of this equation, we can take it that $q_{1} = 0 $ and $p_{1} = 1 $ without any loss of generality. Then the third line of this equation can be rewritten as $\frac{1}{\sqrt{3}} =\sqrt{p_{3}} (\frac{1}{\sqrt{3}}) ^ {n} \alpha_{31}$, which means $\left( \sqrt{3}\right) ^ {n-1} = \sqrt{p_{3}} \alpha_{31}\le \sqrt{p_{3}}\left|\alpha_{31}\right| \le 1$. However, this contradicts to the definiton of cloning: $n>1$. Therefore, $f_{12}=0$.

Now we use our GA to find the optimal probabilistic identification of the following three states which are put in with the same priori probability:
\begin{equation}
\begin{cases}
& \left| \psi_{1}\right\rangle =\frac{1}{\sqrt{3}}(1,1,1) \\
& \left| \psi_{2}\right\rangle =\frac{1}{\sqrt{3}}(1,1,\omega ) ,\omega=e^{\frac{2\pi}{3}i} \\
& \left| \psi_{3}\right\rangle =\frac{1}{\sqrt{3}}(1,\omega,\omega )
\end{cases}
\end{equation}
 we can know from the symmetry of the surface $S$ and the auxiliary plane that $q_{1}=q_{2}=q_{3}$ at their tangent point. Set $q_{1}=q_{2}=q_{3}=q$, and thus, according to formula (\ref{num}),
\begin{equation}
q^{3}-q+\frac{1}{3}=0
\end{equation}
This equation has three roots: $\frac{2\cos\ang{50}}{\sqrt{3}} $,$\frac{2\cos\ang{70}}{\sqrt{3}} $,$\frac{2\cos\ang{170}}{\sqrt{3}} $. The third root is negative and thus is not the one we want. Since the second root is smaller than the first root, the point related to the second root is out of the closure of set $B$ and the first root is the one we want. Therefore, $q=\frac{2\cos\ang{50}}{\sqrt{3}}$. Consequently,
\begin{equation}
Q=3\times \frac{1}{3}q=q=\frac{2\cos\ang{50}}{\sqrt{3}}\approx 0.742
\end{equation}
In other words, the average probability of success of optimal identification of these three states with the priori probabilities $\eta=(\frac{1}{3},\frac{1}{3},\frac{1}{3}) $ is $1-\frac{2\cos\ang{50}}{\sqrt{3}}\approx 0.258$.

\end{document}